\documentstyle[aps,preprint]{revtex}

\draft
\tighten
\begin{document}
\title{Effect of Spin-Flip Scattering on Electrical Transport in Magnetic Tunnel
Junctions}
\author{Zhen-Gang Zhu, Gang Su$^{\ast }$, Qing-Rong Zheng and Biao Jin}
\address{Department of Physics, The Graduate School of the Chinese Academy of\\
Sciences, P.O. Box 3908, Beijing 100039, China}
\maketitle

\begin{abstract}
By means of the nonequilibrium Green function technique, the effect of
spin-flip scatterings on the spin-dependent electrical transport in
ferromagnet-insulator-ferromagnet (FM-I-FM) tunnel junctions is
investigated. It is shown that Julli\`{e}re$^{\prime }$s formula for the
tunnel conductance must be modified when including the contribution from the
spin-flip scatterings. It is found that the spin-flip scatterings could lead
to an angular shift of the tunnel conductance, giving rise to the junction
resistance not being the largest when the orientations of magnetizations in
the two FM electrodes are antiparallel, which may offer an alternative
explanation for such a phenomenon observed previously in experiments in some
FM-I-FM junctions. The spin-flip assisted tunneling is also observed.
\end{abstract}


Spin-dependent electrical transport in magnetic tunnel junctions has
received much attention both theoretically and experimentally in recent
years (see e.g. Refs.\cite{wolf} for review). A new field dubbed as {\it %
spintronics} (i.e. spin-based electronics) is emerging. Among others one of
the simplest devices in spintronics would be a
ferromagnet-insulator-ferromagnet (FM-I-FM) structure which is comprised of
two ferromagnetic electrodes separated by an insulator thin film. In 1975
Julli\`{e}re made the first observation of spin-polarized electrons
tunneling through an insulator film into a ferromagnetic metal film, and
clearly observed 14\% tunnel magnetoresistance (TMR) for Fe/Ge/Co junctions
at 4.2 K\cite{julliere}. In 1995, Moodera et al made a breakthrough that
they observed over a 10\% TMR for a Co/Al$_{2}$O$_{3}$/Ni$_{80}$Fe$_{20}$
junction reproducibly at room temperature \cite{moodera1}. Since then, there
are a variety of works toward enhancing TMR in magnetic tunnel junctions,
and the TMR $>30\%$ has been obtained at room temperature\cite{wolf}. On the
other hand, to understand the spin-polarized tunneling results for FM-I-FM
junctions, people usually invoke the model based on a classical tunneling
theory proposed by Julli\`{e}re\cite{julliere}, in which spins of electrons
during tunneling are supposed to be conserved, namely, the tunneling of
spin-up and spin-down electrons are two quite independent processes, and
spin-flip scatterings are neglected. Though Julli\`{e}re's two-current model
can interpret well some experimental results qualitatively, it still faces
to difficulties for more complex situations. Actually, in some experiments
spin conservation no longer holds, and the spin-flip scattering may take
effect on the transport properties. There has been a number of experiments 
\cite{wolf,experiments} showing that TMR can be very various for different
barriers, and the inverse TMR can even occur, namely, the resistance when
the orientations of magnetizations of the two ferromagnets are parallel, is
larger than that of antiparallel orientations. It appears that the spin-flip
scatterings might not be ignored in these situations. Recently, Vedyayev et
al investigated a model including impurities in the middle of the barrier,
and considered both cases of spin conserving scattering and spin-flip
scattering\cite{vedyayev}. Besides, Jaffr\`{e}s et al measured the angular
dependence of the TMR for transition-metal based junctions, and observed
that the angular response is beyond the simple cosine shape\cite{jaffres}.

In this paper, based on a microscopic model and using the nonequilibrium
Green function technique we shall give a more general expression of the
angular dependence of the TMR for FM-I-FM junctions by including the effect
of spin-flip scatterings. It is found that the effect of spin-flip
scatterings gives rise to a correction to the formula of the usual tunnel
conductance. Specifically, we have found that the spin-flip scattering
induces a phase shift, leading to the tunnel conductance not to be the
smallest when the magnetizations of the two FM electrodes are antiparallel,
which may provide an alternative explanation for the previously experimental
observation of the angular shift in some FM-I-FM tunnel junctions. In
addition, it has been shown that the spin-flip scattering could also lead to
the inverse TMR effect under certain conditions though it is not the only
factor, and the spin-flip assisted tunneling is also observed.

Let us consider a magnetic tunnel junction consisting of two ferromagnetic
films separated by an insulator thin film. A steady bias voltage is applied
to the junction. The relative orientation of magnetizations in the two
ferromagnets is characterized by the angle $\theta $. The Hamiltonian of the
system reads

\begin{equation}
H=H_{L}+H_{R}+H_{T},  \label{Thamiltonian}
\end{equation}
with 
\begin{eqnarray*}
H_{L} &=&\sum_{k\sigma }\varepsilon _{k\sigma }^{L}a_{k\sigma }^{\dagger
}a_{k\sigma }, \\
H_{R} &=&\sum_{q\sigma }[(\varepsilon _{R}({\bf q)-\sigma }M_{2}\cos \theta
)c_{q\sigma }^{\dagger }c_{q\sigma }-M_{2}\sin \theta c_{q\sigma }^{\dagger
}c_{q\overline{\sigma }}], \\
H_{T} &=&\sum_{kq\sigma \sigma ^{\prime }}[T_{kq}^{\sigma \sigma ^{\prime
}}a_{k\sigma }^{\dagger }c_{q\sigma ^{\prime }}+T_{kq}^{\sigma \sigma
^{\prime }}{}^{\ast }c_{q\sigma ^{\prime }}^{\dagger }a_{k\sigma }],
\end{eqnarray*}
where $a_{k\sigma }$ and $c_{k\sigma }$ are annihilation operators of
electrons with momentum $k$ and spin $\sigma $ $(=\pm 1)$ in the left and
right ferromagnets, respectively, $\varepsilon _{k\sigma }^{L}=\varepsilon
_{L}({\bf k)-}eV{\bf -\sigma }M_{1},$ $M_{1}=\frac{g\mu _{B}h_{L}}{2},$ $%
M_{2}=\frac{g\mu _{B}h_{R}}{2},$ $g$ is Land\'{e} factor, $\mu _{B}$ is Bohr
magneton, $h_{L(R)}$ is the molecular field of the left (right) ferromagnet, 
$\varepsilon _{L(R)}({\bf k)}$ is the single-particle dispersion of the left
(right) FM electrode, $V$ is the applied bias voltage, $T_{kq}^{\sigma
\sigma ^{\prime }}$ denotes the spin and momentum dependent tunneling
amplitude through the insulating barrier. Note that the spin-flip scattering
is included in $H_{T}$ when $\sigma ^{\prime }=\bar{\sigma}=-\sigma $. It is
this term that violates the spin conservation in the tunneling process. By
performing the $u-v$ transformation, $c_{q\sigma }=\cos \frac{\theta }{2}%
b_{q\sigma }-\sigma \sin \frac{\theta }{2}b_{q\overline{\sigma }},$ $%
c_{q\sigma }^{\dagger }=\cos \frac{\theta }{2}b_{q\sigma }^{\dagger }-\sigma
\sin \frac{\theta }{2}b_{q\overline{\sigma }}^{\dagger },$ $H_{R}$ becomes $%
H_{R}=\sum_{q\sigma }\varepsilon _{q\sigma }^{R}b_{q\sigma }^{\dagger
}b_{q\sigma },$ with $\varepsilon _{q\sigma }^{R}=\varepsilon _{R}({\bf %
q)-\sigma }M_{2},$ and $H_{T}$ becomes $H_{T}=\sum_{kq\sigma \sigma ^{\prime
}}T_{kq}^{\sigma \sigma ^{\prime }}(\cos \frac{\theta }{2}a_{k\sigma
}^{\dagger }b_{q\sigma ^{\prime }}-\sigma ^{\prime }\sin \frac{\theta }{2}%
a_{k\sigma }^{\dagger }b_{q\overline{\sigma }^{\prime }})+h.c..$ The
tunneling current has the form of

\begin{equation}
I_{L}(V)=e\left\langle \stackrel{\cdot }{N_{L}}\right\rangle =-\frac{2e}{%
\hbar }%
%TCIMACRO{\func{Re}}%
%BeginExpansion
\mathop{\rm Re}%
%EndExpansion
\sum_{kq}Tr_{\sigma }\{{\bf \Omega }_{kq}{\bf \cdot G}_{kq}^{<}(t,\text{ }%
t)\},  \label{c}
\end{equation}
where ${\bf \Omega }_{kq}{\bf =T}_{kq}\cdot {\bf R}$ with ${\bf T}%
_{kq}=\left( 
\begin{array}{cc}
T_{kq}^{\uparrow \uparrow } & T_{kq}^{\uparrow \downarrow } \\ 
T_{kq}^{\downarrow \uparrow } & T_{kq}^{\downarrow \downarrow }
\end{array}
\right) $ and ${\bf R=}\left( 
\begin{array}{cc}
\cos \frac{\theta }{2} & -\sin \frac{\theta }{2} \\ 
\sin \frac{\theta }{2} & \cos \frac{\theta }{2}
\end{array}
\right) ,$ and $Tr_{\sigma }$ stands for the trace of matrix taking over the
spin space. The lesser Green function, ${\bf G}_{kq}^{<}(t,$ $t^{\prime }),$
is defined in a steady state as

\begin{equation}
{\bf G}_{kq}^{<}(t-\text{ }t^{\prime })=\left( 
\begin{array}{cc}
G_{kq}^{\uparrow \uparrow ,<}(t-t^{\prime }) & G_{kq}^{\downarrow \uparrow
,<}(t-t^{\prime }) \\ 
G_{kq}^{\uparrow \downarrow ,<}(t-t^{\prime }) & G_{kq}^{\downarrow
\downarrow ,<}(t-t^{\prime })
\end{array}
\right) ,  \label{glesser}
\end{equation}
with $G_{kq}^{\sigma \sigma ^{\prime },<}(t-t^{\prime })\equiv i\left\langle
a_{k\sigma }^{\dagger }(t^{\prime })b_{q\sigma ^{\prime }}(t)\right\rangle .$
To obtain the lesser Green function, one needs to introduce a time-ordered
Green function ${\bf G}_{qk}^{t}$ as 
\begin{equation}
{\bf G}_{qk}^{t}(t-t^{\prime })=\left( 
\begin{array}{cc}
G_{qk}^{\uparrow \uparrow ,t}(t-t^{\prime }) & G_{qk}^{\uparrow \downarrow
,t}(t-t^{\prime }) \\ 
G_{qk}^{\downarrow \uparrow ,t}(t-t^{\prime }) & G_{qk}^{\downarrow
\downarrow ,t}(t-t^{\prime })
\end{array}
\right) ,  \label{firstGF}
\end{equation}
with $G_{qk}^{\sigma \sigma ^{\prime },t}(t-t^{\prime })\equiv
-i\left\langle T\{b_{q\sigma }(t)a_{k\sigma ^{\prime }}^{\dagger }(t^{\prime
})\}\right\rangle .$ By using the equation of motion, we get

\begin{equation}
{\bf G}_{qk}^{t}(\varepsilon )=\sum_{q^{\prime }}{\bf F}_{qq^{\prime
}}^{t}(\varepsilon ){\bf \Omega }_{kq^{\prime }}^{\dagger }{\bf g}%
_{kL}^{t}(\varepsilon ),  \label{green1}
\end{equation}
with 
\begin{equation}
{\bf F}_{qq^{\prime }}^{t}(\varepsilon )={\bf g}_{q^{\prime
}R}^{t}(\varepsilon )\delta _{qq^{\prime }}+\sum_{k^{\prime }}{\bf G}%
_{qk^{\prime }}^{t}(\varepsilon ){\bf \Omega }_{k^{\prime }q^{\prime }}{\bf g%
}_{q^{\prime }R}^{t}(\varepsilon ),  \label{fdyson}
\end{equation}
where the use has been made of the Fourier transform of the time-ordered
Green function ${\bf G}_{qk}^{t}(\varepsilon )=\int dte^{i\varepsilon
(t-t^{\prime })}{\bf G}_{qk}^{t}(t-t^{\prime }),$ and ${\bf g}%
_{kL}^{t}(\varepsilon )$, ${\bf g}_{q^{\prime }R}^{t}(\varepsilon )$ are the
time-ordered Green function of the left and right FM electrodes for the
uncoupled system, respectively. By applying the Langreth theorem \cite{haug}
to Eq.(\ref{green1}), we get

\begin{equation}
{\bf G}_{kq}^{<}(\varepsilon )=\sum_{q^{\prime }}[{\bf F}_{qq^{\prime
}}^{r}(\varepsilon ){\bf \Omega }_{kq^{\prime }}^{\dagger }{\bf g}%
_{kL}^{<}(\varepsilon )+{\bf F}_{qq^{\prime }}^{<}(\varepsilon ){\bf \Omega }%
_{kq^{\prime }}^{\dagger }{\bf g}_{kL}^{a}(\varepsilon )],  \label{lesserGF}
\end{equation}
where the superscript ``r(a)'' denotes the ``retarded (advanced )'' Green
function. The tunneling current becomes 
\begin{equation}
I_{L}(V)=-\frac{2e}{\hbar }\int \frac{d\varepsilon }{2\pi }%
%TCIMACRO{\func{Re}}%
%BeginExpansion
\mathop{\rm Re}%
%EndExpansion
(\sum_{k,q,q^{\prime }}\{Tr_{\sigma }({\bf \Omega }_{kq}[{\bf F}_{qq^{\prime
}}^{r}(\varepsilon ){\bf \Omega }_{kq^{\prime }}^{\dagger }{\bf g}%
_{kL}^{<}(\varepsilon )+{\bf F}_{qq^{\prime }}^{<}(\varepsilon ){\bf \Omega }%
_{kq^{\prime }}^{\dagger }{\bf g}_{kL}^{a}(\varepsilon )])\}).
\label{newcurrent}
\end{equation}
To get useful analytical result we may assume for simplicity that the
tunneling amplitude ${\bf T}_{kq}$ is independent of the momentum like the
conventional consideration\cite{wolf}, but depends on spin. This suggests
that apart from inclusion of the spin-flip scattering we have supposed that
the tunneling amplitude of electrons for the spin-up channel differs from
that of the spin-down channel. As a result, ${\bf T}_{kq}$ becomes ${\bf T}%
=( 
\begin{array}{cc}
T_{1} & T_{2} \\ 
T_{3} & T_{4}
\end{array}
).$ In addition, the elements of ${\bf T}$ are assumed to be real. Up to the
first-order approximation to the Green function ${\bf F}_{qq^{\prime
}}^{t}(\varepsilon )$, we get

\begin{equation}
I_{L}(V)=\frac{2e}{\hbar }%
%TCIMACRO{\func{Re}}%
%BeginExpansion
\mathop{\rm Re}%
%EndExpansion
\int \frac{d\varepsilon }{2\pi }[f(\varepsilon )-f(\varepsilon
+eV)]Tr_{\sigma }[{\bf T}_{eff}(\varepsilon ,V)],  \label{currentall}
\end{equation}
where $f(\varepsilon )$ is the Fermi function,

\begin{equation}
{\bf T}_{eff}(\varepsilon ,V)=2\pi ^{2}{\bf T\cdot R\cdot D}_{R}(\varepsilon
)\cdot {\bf R}^{\dagger }\cdot {\bf T}^{\dagger }\cdot {\bf D}%
_{L}(\varepsilon +eV),  \label{transmission}
\end{equation}
and ${\bf D}_{L(R)}(\varepsilon )$ is a $2\times 2$ diagonal matrix with two
nonzero elements being the corresponding density of states (DOS) of
electrons with spin up and down in the left (right) ferromagnet. For a small
bias voltage $V,$ we obtain the tunneling conductance 
\begin{equation}
G=\frac{2e^{2}}{h}T_{eff},  \label{G-butti}
\end{equation}
where $T_{eff}=Tr_{\sigma }[%
%TCIMACRO{\func{Re}}%
%BeginExpansion
\mathop{\rm Re}%
%EndExpansion
({\bf T}_{eff}(\varepsilon _{F},V=0))]$, and $\varepsilon _{F}$ is the Fermi
energy. In comparison to the Landauer-B\"{u}ttiker formula, one may find
that $T_{eff}$ can be regarded as an effective tunneling transmission
coefficient which includes the contribution from spin-flip scatterings. Eq. (%
\ref{G-butti}) can be explicitly rewritten as

\begin{equation}
G=G_{0}[1+P_{2}\sqrt{P_{1}^{2}+P_{3}^{2}}\cos (\theta -\theta _{f})],
\label{G}
\end{equation}
where $G_{0}=\frac{\pi e^{2}}{2\hbar }[(T_{1}^{2}+T_{2}^{2})D_{L\uparrow
}+(T_{3}^{2}+T_{4}^{2})D_{L\downarrow }](D_{R\uparrow }+D_{R\downarrow })$, $%
P_{1}=\frac{(T_{1}^{2}-T_{2}^{2})D_{L\uparrow
}-(T_{4}^{2}-T_{3}^{2})D_{L\downarrow }}{(T_{1}^{2}+T_{2}^{2})D_{L\uparrow
}+(T_{3}^{2}+T_{4}^{2})D_{L\downarrow }},$ $P_{2}=\frac{D_{R\uparrow
}-D_{R\downarrow }}{D_{R\uparrow }+D_{R\downarrow }},$ $P_{3}=\frac{%
2(T_{1}T_{2}D_{L\uparrow }+T_{3}T_{4}D_{L\downarrow })}{%
(T_{1}^{2}+T_{2}^{2})D_{L\uparrow }+(T_{3}^{2}+T_{4}^{2})D_{L\downarrow }},$ 
$\tan \theta _{f}=\frac{P_{3}}{P_{1}},$ $D_{L\uparrow }=D_{L}(\varepsilon
+M_{1}+eV),$ $D_{L\downarrow }=D_{L}(\varepsilon -M_{1}+eV),$ $D_{R\uparrow
}=D_{R}(\varepsilon +M_{2}),$ $D_{R\downarrow }=D_{R}(\varepsilon -M_{2}),$
and $D_{L(R)}$ is the DOS of electrons in the left (right) ferromagnet. One
may observe that there is an angular shift induced by the spin-flip
scatterings, as to be discussed below.

Now let us look at the angular dependence of the conductance $G$. When the
spin-flip scattering is neglected, i.e. $T_{2}=T_{3}=0$, and if we further
assume $T_{1}=T_{4}$, we recover the conventional expression for the
conductance $G=\bar{G}_{0}(1+\bar{P}_{1}P_{2}\cos \theta ),$ which is
familiar in literature\cite{wolf,slon}, where $\bar{G}_{0}=\frac{\pi e^{2}}{%
2\hbar }T_{1}^{2}(D_{L\uparrow }+D_{L\downarrow })(D_{R\uparrow
}+D_{R\downarrow })$, and $\bar{P}_{1}$ $=\frac{D_{L\uparrow
}-D_{L\downarrow }}{D_{L\uparrow }+D_{L\downarrow }}$ is the usual
polarization of the left ferromagnet, and $P_{3}=0.$ When $T_{2}=T_{3}=0$
but $T_{1}\neq T_{4}$, which implies that even if the spin-flip scattering
is ignored, but the tunneling amplitude of electrons for the spin-up channel
is different from that for the spin-down channel, we can also get in this
situation an expression 
\begin{equation}
G=G_{0}^{\prime }(1+P_{1}^{\prime }P_{2}\cos \theta ),  \label{another}
\end{equation}
where $G_{0}^{\prime }=\frac{\pi e^{2}}{2\hbar }(T_{1}^{2}D_{L\uparrow
}+T_{4}^{2}D_{L\downarrow })(D_{R\uparrow }+D_{R\downarrow })$ and $%
P_{1}^{\prime }=\frac{T_{1}^{2}D_{L\uparrow }-T_{4}^{2}D_{L\downarrow }}{%
T_{1}^{2}D_{L\uparrow }+T_{4}^{2}D_{L\downarrow }}.$ Although it looks
seemingly like the conventional form, it is clear that the difference of the
tunneling amplitudes for the two independent spin channels can still alter
the magnitude of the conductance and the polarization as well.

On the other hand, the angular dependence of the tunnel conductance without
considering the spin-flip effects, as mentioned before, is well known: 
\begin{equation}
G=G_{P}\cos ^{2}\frac{\theta }{2}+G_{AP}\sin ^{2}\frac{\theta }{2},
\label{G-conven}
\end{equation}
with $G_{P}$ the conductance for parallel orientation of magnetizations in
the two FM electrodes, and $G_{AP}$ the conductance for the antiparallel
orientation (see e.g. Refs.\cite{wolf,jaffres}). While in the present case,
namely, with inclusion of the effect of spin-flip scatterings we find from
Eq.(\ref{G}) that the angular dependence of the conductance becomes

\begin{equation}
G=G_{1}\cos ^{2}\frac{\theta }{2}+G_{2}\sin ^{2}\frac{\theta }{2}+G_{3}\sin
\theta ,  \label{G-flip}
\end{equation}
where $G_{1}=\frac{\pi e^{2}}{2\hbar }\{D_{R\uparrow }[T_{1}^{2}D_{L\uparrow
}+T_{3}^{2}D_{L\downarrow }]+D_{R\downarrow }[T_{2}^{2}D_{L\uparrow
}+T_{4}^{2}D_{L\downarrow }]\}$, $G_{2}=\frac{\pi e^{2}}{2\hbar }%
\{D_{R\uparrow }[T_{2}^{2}D_{L\uparrow }+T_{4}^{2}D_{L\downarrow
}]+D_{R\downarrow }[T_{1}^{2}D_{L\uparrow }+T_{3}^{2}D_{L\downarrow }]\},$
and $G_{3}=\frac{\pi e^{2}}{2\hbar }(D_{R\uparrow }-D_{R\downarrow
})(T_{1}T_{2}D_{L\uparrow }+T_{3}T_{4}D_{L\downarrow })$. One may see that
apart from the conventional $\cos ^{2}\frac{\theta }{2}$ and $\sin ^{2}\frac{%
\theta }{2}$ terms there is an additional third term proportional to $\sin
\theta $. Here we should point out that $G_{1}$ is the conductance in the
case of the parallel alignment ($\theta =0$) for the magnetizations of the
two ferromagnets, $G_{2}$ is corresponding to the antiparallel case ($\theta
=\pi $), and $G_{3}$ gives an additional term for the noncollinear case ($%
\theta \neq 0$ and $\pi $), which disappears in the collinear cases.
Certainly, these three coefficients $G_{i}$ ($i=1,2,3$) contain the
contributions from the spin-flip scatterings characterized by $T_{2}$ and $%
T_{3}$. It is the effect of spin-flip scatterings that enables the tunnel
conductance not to be at the minimum when the magnetizations of the two
ferromagnets are antiparallel. This is understandable, because the spin-flip
scattering process violates the spin conservation and can enable electrons
in the spin-up band of one FM electrode tunneling through the insulator
barrier into the spin-down band of another FM electrode, and vice versa,
thereby giving rise to a phase shift, as shown in Eq.(\ref{G}). It is
emphasized that this shift will disappear when the effect of spin-flip
scattering is neglected. Therefore, Eqs. (\ref{G}) and (\ref{G-flip}) can be
viewd as a generalization of the conventional expression for the tunnel
conductance [see Eq.(\ref{G-conven})]. It is interesting to note that the
phenomenon of such an angular shift has been experimentally observed for a
CoFe/Al$_{2}$O$_{3}$/Co tunnel junction, as presented in Ref. \cite{moodera2}
(see Fig.4 therein), where the maximum of the junction resistance appears at 
$\theta =200^{\circ }$, not $180^{\circ }$, implying the angular shift $%
\theta _{f}$ $=20^{\circ }$. Although the authors of Ref.\cite{moodera2} did
not mention the reasons why such an angular shift occurs in this FM-I-FM
junction, in accordance with the aforementioned analysis we may attribute
this phenomenon possibly to the effect of the spin-flip scatterings. If this
is acceptable, we can in turn infer the magnitude of the effect of spin-flip
scatterings. To show it explicitly, let us assume $T_{1}\approx T_{4}$ and $%
T_{2}\approx T_{3}$ for simplicity, and \hspace{0in}define a parameter 
\begin{equation}
\gamma =\frac{T_{2}}{T_{1}},
\end{equation}
which characterizes the magnitude of the effect of spin-flip scatterings.
The angular shift $\theta _{f}$ versus the parameter $\gamma $ is plotted in
Fig. 1. It is seen that $\theta _{f}$ is monotonously increasing with
increasing $\gamma .$ When $\gamma $ approaches to $1$, $\theta _{f}$ $%
=90^{\circ }$ which can be obtained from the expression of $P_{1}$ (see
below). If $\gamma >1$, then $P_{1}$ can be negative, leading to $\theta
_{f} $ larger than $90^{\circ }$. To see more clearly the effect of
spin-flip scatterings on the conductance, we note that $G_{0}$ in Eq. (\ref
{G}) can be written as $G_{0}=(1+\gamma ^{2})\bar{G}_{0}$. In this case, $%
P_{1}=\frac{1-\gamma ^{2}}{1+\gamma ^{2}}\bar{P}_{1}$, and $P_{3}=\frac{%
2\gamma }{1+\gamma ^{2}}$. The $\gamma $-dependence of the conductance is
presented in Fig. 2 (a). \ It can be seen that the conductance decreases
with increasing $\gamma $ when the magnetizations in the two ferromagnets
are parallel, while it increases for the antiparallel alignment. As $\gamma
>1$, one may observe that $G(\theta =\pi )>G(\theta =0)$, suggesting that
the inverse TMR may occur. For the case of noncollinear alignments, with
increasing $\gamma $ the conductance first increases rapidly and then
decreases, and some peaks appear around $\gamma \approx 1$.

We come to consider the effect of spin-flip scatterings on the TMR.
Recently, a number of experiments\cite{experiments} for a few FM-I-FM tunnel
junctions show that if the insulator thin film is the material which differs
from Al$_{2}$O$_{3},$ such as SrTiO$_{3}$, Ce$_{0.69}$La$_{0.31}$O$_{1.845}$
and so on, the TMR, defined as usual as $1-G(\theta =\pi )/G(\theta =0),$
will be negative under certain conditions, which means $G(\theta =\pi
)>G(\theta =0),$ exhibiting the so-called inverse TMR effect. It is
generally believed that this effect may originate from the electronic states
at the interface between a ferromagnetic layer and an insulating layer\cite
{experiments} which could give rise to the density of states in the minority
spin band larger than that in the majority spin band at the Fermi level.
However, one may see below that the spin-flip scattering can also contribute
to the inverse TMR. From (\ref{G}) we find that the TMR still has the
apparently standard form 
\begin{equation}
TMR=\frac{2P_{1}P_{2}}{1+P_{1}P_{2}},  \label{Tmr}
\end{equation}
but $P_{1}$, defined after Eq. (\ref{G}) and containing the contribution
from spin-flip scatterings, differs from the conventional polarization $\bar{%
P}_{1}$. In Ref. \cite{moodera1}, Moodera et al calculated the TMR for the
CoFe/Al$_{2}$O$_{3}$/Co junction according to the Julli\`{e}re$^{\prime }$s
formula. The calculated result is $27\%$, while the experiment value is $24\%
$ at $4.2$ $K$. If we adopt this experimental data, we can calculate the
contribution of the spin-flip scatterings which is characterized by the
parameter $\gamma $. The obtained result for $\gamma $ is about $0.28$,
where the spin polarization of electrons is taken as $47\%$ for CoFe and $%
34\%$ for Co\cite{moodera1}. It shows that the spin-flip scatterings might
have a considerable effect on the electrical transport of this tunnel
junction. On the other hand, if we take the contribution from the spin-flip
scatterings into account, then we can apply our formula to estimate the
value of TMR, which could be closer to the experimental result than using
Julli\`{e}re$^{\prime }$s formula. If the tunneling amplitudes satisfy a
certain condition, $P_{1}$ can be negative, depending on the difference
between $(T_{1}^{2}-T_{2}^{2})D_{L\uparrow }$ and $(T_{4}^{2}-T_{3}^{2})D_{L%
\downarrow }$, resulting in that the TMR can be negative. The $\gamma $%
-dependence of the TMR is depicted in Fig. 2 (b). It can be seen that the
TMR decreases with increasing $\gamma $, and becomes negative for $\gamma >1$%
. This can be understood in the following. For $0<\gamma <1$, the tunneling
amplitude for the two independent channels ($T_{1}$) is larger than the
tunneling amplitude for the spin-flip channel ($T_{2}$). Owing to the
spin-flip scatterings the polarization $P_{1}$ becomes effective and small,
leading to decreasing of the TMR. When $\gamma >1$, i.e., the tunneling
amplitude for the two independent channels is smaller than that for the
spin-flip channel, the spin-flip scattering dominates in the tunneling
process, implying that the electrons with spin up in the left FM electrode
can tunnel through the insulator barrier to occupy the states of electrons
with spin down in the right FM electrode via the spin-flip mechanism,
thereby giving rise to contribution to the inverse TMR effect. 

The angular dependence of the TMR can be understood from the following
definition

\begin{equation}
TMR(\theta )=\frac{G(\theta )-G(\theta =\pi )}{G(\theta =0)}=\frac{%
P_{2}(P_{1}+\sqrt{P_{1}^{2}+P_{3}^{2}}\cos (\theta -\theta _{f}))}{%
1+P_{1}P_{2}}.  \label{TmrTheta}
\end{equation}
When $\theta $ $=$ $\theta _{f}$, the TMR goes to its maximum which will be
denoted by $TMR(\theta _{f})$ hereafter. In Fig. 3, the $\gamma $-dependence
of the maximum TMR, i.e. $TMR(\theta _{f})$, is presented. A remarkable
property is that $TMR(\theta _{f})$ has a peak at $\gamma $ $\approx $ $0.82$%
, which migh be a result of the spin-flip assisted tunneling. It is seen
that $TMR(\theta _{f})$ approaches to a constant when $\gamma $ increases to
a large value. No matter how large $\gamma $ is, $TMR(\theta _{f})$ is
always positive.

The above discussion is based on the result up to the first-order
approximation for the Green functions. To include contributions from
higher-order Green functions, it is better to consider them in a Keldysh
space. We introduce the nonequilibrium Green function in the Keldysh space
as \cite{haug}

\begin{equation}
\widehat{{\bf G}}_{qk}(t,\text{ }t^{\prime })=\left( 
\begin{array}{cc}
{\bf G}_{qk}^{t}(t,\text{ }t^{\prime }) & {\bf G}_{kq}^{<}(t,\text{ }%
t^{\prime }) \\ 
{\bf G}_{qk}^{>}(t,\text{ }t^{\prime }) & {\bf G}_{qk}^{\widetilde{t}}(t,%
\text{ }t^{\prime })
\end{array}
\right) ,  \label{KeldyshGf}
\end{equation}
where ${\bf G}_{qk}^{t}(t,$ $t^{\prime })$ is the time-ordered Green
function defined as before, ${\bf G}_{kq}^{<(>)}(t,$ $t^{\prime })$ is
lesser (greater) Green function, and ${\bf G}_{qk}^{\widetilde{t}}(t,$ $%
t^{\prime })$ is the antitime-ordered Green function. The elements of ${\bf G%
}_{qk}^{\widetilde{t}}(t,$ $t^{\prime })$ and ${\bf G}_{qk}^{>}(t,$ $%
t^{\prime })$ are given by

\begin{eqnarray}
G_{qk}^{\sigma \sigma ^{\prime },\widetilde{t}}(t,\text{ }t^{\prime })
&=&-i\langle \widetilde{T}[b_{q\sigma }(t)a_{k\sigma ^{\prime }}^{\dagger
}(t^{\prime })]\rangle ,  \label{defineGf} \\
G_{kq}^{\sigma \sigma ^{\prime },>}(t,\text{ }t^{\prime }) &=&-i\left\langle
b_{q\sigma }(t)a_{k\sigma ^{\prime }}^{\dagger }(t^{\prime })\right\rangle .
\nonumber
\end{eqnarray}
According to Eqs. (\ref{green1}) and (\ref{fdyson}), we can make a sum of
Feynman diagrams shown in Fig. 4. The result is 
\begin{equation}
\widehat{{\bf G}}_{kq}(\varepsilon )=\widehat{{\bf g}}_{R}(\varepsilon )%
\widehat{{\bf \Sigma }}(\varepsilon )\widehat{{\bf g}}_{L}(\varepsilon ),
\label{dyson}
\end{equation}
where $-i\widehat{{\bf \Sigma }}(\varepsilon )=-i\Omega ^{\dagger }\widehat{%
\tau }_{3}(\widehat{1}-\widehat{{\bf \eta }})^{-1}$, $\widehat{{\bf \eta }}=%
\widehat{{\bf g}}_{L}(\varepsilon )\Omega \widehat{\tau }_{3}\widehat{{\bf g}%
}_{R}(\varepsilon )\Omega ^{\dagger }\widehat{\tau }_{3}$, ${\bf \Omega }=%
{\bf T\cdot R}$, and $\widehat{\tau }_{3}$ is a Pauli matrix. From Eq. (\ref
{c}), the current can be rewritten as

\begin{equation}
I_{L}(V)=-\frac{2e}{\hbar }%
%TCIMACRO{\func{Re}}%
%BeginExpansion
\mathop{\rm Re}%
%EndExpansion
\int \frac{d\varepsilon }{2\pi }Tr_{\sigma }\{({\bf \Omega }\widehat{{\bf G}}%
_{qk}(\varepsilon ))_{12}\}.  \label{currentnow}
\end{equation}
After a tedious calculation, we get

\begin{eqnarray}
I_{L}(V) &=&-\frac{e}{\hbar }%
%TCIMACRO{\func{Re}}%
%BeginExpansion
\mathop{\rm Re}%
%EndExpansion
\int \frac{d\varepsilon }{2\pi }Tr\{\widehat{{\bf F}}_{2}(\widehat{\tau }%
_{3}-i\widehat{\tau }_{2})\widehat{{\bf F}}_{1}{\bf T}_{eff}(\varepsilon
,V)[(1+{\bf T}_{eff}(\varepsilon ,V)/2)\widehat{\tau }_{0}  \nonumber \\
&&+(f(\varepsilon +eV)-f(\varepsilon ))(\widehat{\tau }_{3}+i\widehat{\tau }%
_{2}){\bf T}_{eff}(\varepsilon ,V)]^{-1}\},  \label{currentmatrix}
\end{eqnarray}
where $Tr$ denotes the trace over the spin space and the Keldysh space, $%
\widehat{\tau }_{0}$ is the unit matrix, $\widehat{\tau }_{i}$ $(i=1,2,3)$
are Pauli matrices, and 
\[
\widehat{{\bf F}}_{1}=\left( 
\begin{array}{cc}
1-2f(\varepsilon ) & 0 \\ 
0 & 2f(\varepsilon )
\end{array}
\right) \text{; }\widehat{{\bf F}}_{2}=\left( 
\begin{array}{cc}
2f(\varepsilon +eV) & 0 \\ 
0 & 1-2f(\varepsilon +eV)
\end{array}
\right) . 
\]
Eq. (\ref{currentmatrix}) can be rewritten in a compact form

\begin{equation}
I_{L}(V)=\frac{4e}{h}%
%TCIMACRO{\func{Re}}%
%BeginExpansion
\mathop{\rm Re}%
%EndExpansion
\int d\varepsilon \lbrack f(\varepsilon +eV)-f(\varepsilon )]Tr_{\sigma }\{[%
{\bf \Lambda }(\varepsilon ,V)-1]\cdot {\bf \Lambda }(\varepsilon ,V)\},
\label{allsumcurrent}
\end{equation}
where ${\bf \Lambda (}\varepsilon ,V)=\frac{{\bf T}_{eff}(\varepsilon ,V)}{2}%
\cdot (1+\frac{{\bf T}_{eff}(\varepsilon ,V)}{2})^{-1}$ and ${\bf T}%
_{eff}(\varepsilon ,V)=\left( 
\begin{array}{cc}
T_{eff}^{1}(\varepsilon ,V) & T_{eff}^{2}(\varepsilon ,V) \\ 
T_{eff}^{3}(\varepsilon ,V) & T_{eff}^{4}(\varepsilon ,V)
\end{array}
\right) $ are matrices in spin space, and $T_{eff}^{i}(i=1,2,3,4)$ are the
elements of the effective transmission matrix ${\bf T}_{eff}(\varepsilon ,V)$
given in Eq. (\ref{transmission}). In principle, Eq. (\ref{allsumcurrent})
gives the current including more corrections from the spin-flip scatterings,
and the tunnel conductance can be obtained by $G=\partial I_{L}(V)/\partial
V $. For a small bias voltage, we get 
\begin{equation}
G=\frac{2e^{2}}{h}\tilde{T}_{eff}\text{,}  \label{G-allsum}
\end{equation}
where $\tilde{T}_{eff}=\frac{[8T_{eff}^{0}+(T_{eff}^{1}(\varepsilon
_{F})+T_{eff}^{4}(\varepsilon _{F}))(2+T_{eff}^{0})]}{%
[2+T_{eff}^{0}+(T_{eff}^{1}(\varepsilon _{F})+T_{eff}^{4}(\varepsilon
_{F}))]^{2}}$ with $T_{eff}^{0}=2\pi ^{4}D_{L\uparrow }(\varepsilon
_{F})D_{L\downarrow }(\varepsilon _{F})D_{R\uparrow }(\varepsilon
_{F})D_{R\downarrow }(\varepsilon _{F})(T_{1}T_{4}-T_{2}T_{3})^{2}$, can be
viewed as the effective transmission coefficient. Compared to Eq. (\ref
{G-butti}), Eq. (\ref{G-allsum}) includes more corrections from the
spin-flip scatterings. Although the angular dependence of $G(\theta )$
determined by Eq. (\ref{G-allsum}) looks more complex in form than one
presented in Eq. (\ref{G}), the behavior of $G(\theta )$ versus $\theta $ is
found to be qualitatively similar to those shown in Fig. 2 (a) for a given $%
\gamma $.

In summary, we have investigated the effect of spin-flip scatterings on the
spin-dependent electrical transport in ferromagnet-insulator-ferromagnet
tunnel junctions by using the nonequilibrium Green function technique. When
the effect of the spin-flip scatterings is taken into account, the
frequently used Julli\`{e}re$^{\prime }$s formula for the tunnel conductance
must be modified, though the form looks seemingly similar. It is found that
the spin-flip scatterings could lead to an angular shift of the tunnel
conductance, giving rise to the junction resistance not being the largest
when the orientations of magnetizations in the two FM electrodes are
antiparallel, which is quite consistent with the experimental observation in
CoFe/Al$_{2}$O$_{3}$/Co tunnel junctions. As the Julli\`{e}re$^{\prime }$s
formula overestimates the value of the TMR, our derived formula with
inclusion of the effect of spin-flip scatterings could estimate the TMR
value closer to the experimental result, as discussed above. It is found
that the spin-flip scattering could also lead to the inverse TMR effect
under certain conditions, though it is not the only factor. The phenomenon
of spin-flip assisted tunneling is clearly observed. When including
high-order terms of the Green function, the angular dependence of the tunnel
conductance is qualitatively similar, although the form looks more complex.
Finally, we would like to mention that our present derivation can be readily
extended to other magnetic junctions.

\section*{Acknowledgments}

This work is supported in part by the National Science Foundation of China
(Grant No. 90103023, 10104015), the State Key Project for Fundamental
Research in China, and by the Chinese Academy of Sciences.

\newpage

{\bf FIGURE CAPTIONS}

Fig. 1 The angular shift $\theta _{f}$ versus the parameter $\gamma $, where
the mass of single electron in both ferromagnets are assumed as unity, the
molecular fields in the two ferromagnets are supposed to be the same and
taken as $0.7eV$, $\varepsilon _{F}=1.5eV$, and the coupling parameter $%
T_{1} $ is chosen as $0.01$ $eV$.

Fig. 2 The $\gamma $-dependence of the tunnel conductance (a) and the TMR
(b), where the parameters are taken the same as those in Fig. 1.

Fig. 3 The $\gamma $-dependence of the maximum tunnel magnetoresistance $%
TMR(\theta _{f})$, where the parameters are taken the same as those in Fig.
1.

Fig. 4 Feynman diagrams for the Green function $\widehat{{\bf G}}%
_{qk}(\varepsilon )$.

\end{document}